\documentclass[prd,aps]{revtex4}

\usepackage{graphicx,subfigure}

\newcommand{\be}{\begin{equation}}
\newcommand{\ee}{\end{equation}}
\newcommand{\bea}{\begin{eqnarray}}
\newcommand{\eea}{\end{eqnarray}}
\newcommand{\bd}{\begin{displaymath}}
\newcommand{\ed}{\end{displaymath}}

\newcommand{\ad }{a^{\dagger}}

\newcommand{\xh }{ \hat{x}}
\newcommand{\ph }{ \hat{p}}
\newcommand{\hh }{ \hat{H}}
\newcommand{\lb }{ \left (}
\newcommand{\rb }{ \right )}

\begin{document}

%\draft
%\preprint{
%\begin{tabular}{l}
%\hbox to\hsize{\hfill KAIST-TH 2006/10}\\
%[-1mm]
%\hbox to\hsize{\hfill KIAS-P06047}\\
%[-2mm] \hbox to\hsize{\hfill hep-ph/yymmdd}\\
%[-3mm] \hbox to\hsize{\hfill November 2011}\\
%[-3mm]
%\end{tabular}
%}

\title{
On the quantization and generalized uncertainty relation for some quantum deformed algebras }

\author{ Won Sang Chung }
\email{mimip4444@hanmail.net}

\affiliation{
Department of Physics and Research Institute of Natural Science, College of Natural Science, Gyeongsang National University, Jinju 660-701, Korea
}

\date{\today}

\begin{abstract}
In this paper, the quantization and generalized uncertainty relation for some quantum deformed algebras are investigated. For several deformed algebras, the commutation relation between the position and the momentum operator is shown to be the same as the one in the Planck scale deformed quantum mechanics. 
\end{abstract}

%\pacs{PACS numbers:12.60.Fr,12.60.Cn,14.80.Cp}
\maketitle

\section{Introduction}

In the last decade quantum groups have been the subject of intensive research in several physics and mathematics fields. From the mathematical point of view, it is a non-commutative associative Hopf algebra. The structure and representation theory of quantum groups have been developed extensively by Jimbo [1] and Drinfeld [2]. The notion of quantum group is easily approached through the one of quantum algebra, which corresponds to a deformation, depending on a certain parameter, of a Lie algebra.

The classical oscillator algebra is defined by the canonical commutation relations :
\be
[a, \ad ] =1, ~~~ [N, a]=-a, ~~ ~ [ N,\ad ]= \ad ,
\ee
where $N$ is called a number operator and it is assumed to be hermitian.

It allows for the different  types of deformations [3,4,5,6,7]. The first deformation was accomplished by Arik and Coon [3]. They used the q-calculus which was originally introduced by Jackson in the early 20th century [8]. In the study of the basic hypergeometric function Jackson invented the Jackson derivative and integral, which is now called q-derivative and q-integral. Jackson's pioneering research enabled theoretical physicists and mathematician to study the new physics or mathematics related to the q-calculus. Much was accomplished in this direction and work is under way to find the meaning of the deformed theory.

Arik and Coon's q-oscillator algebra is given by
\be
aa^{\dagger} - q a^{\dagger} a =1 , ~~~
[ N, a^{\dagger} ] = a^{\dagger}, ~~ [N, a]= -a,
\ee
where the relation between the number operator and step operators becomes
\be
\ad a = [N]_q ,
\ee
where a q-number is defined as
\be
[X]_q = \frac{1-q^X }{1-q }
\ee
for any operator or number $X$.
The algebra (2) was shown to be related to the deformation of the uncertainty relation which is called a generalized uncertainty relation suggested by Kempf [9]. It is thought that the generalized uncertainty relation results from the quantum gravity effect. In Kempf's another work [10], the generalized uncertainty relation was shown to lead to the conclusion that anomalies observed with fields over unsharp coordinates might be testable if the onset of strong gravity effects is not too far above the currently experimentally accessible scale about $ 10^{-18}  $m, rather than at the Planck scale of $10^{-35} $m. The quantum mechanics related to the generalized uncertainty relation is called a Planck scale deformed quantum mechanics, which is regarded as as reasonable model for studying the quantum gravity phenomenology [11,12,13,14,15].

The second deformation was achieved by Macfarlane [4] and Biedenharn [5]. Their q-oscillator algebra is given by
\be
aa^{\dagger} - q a^{\dagger}a=q^{-N},~~~
[ N, a^{\dagger} ] = a^{\dagger}, ~~ [N, a]= -a,
\ee
where the relation between the number operator and step operators becomes
\be
a^{\dagger} a = q^{ 1-N  }[N]_q ,
\ee
The third deformation is obtained by Chung {\it et.al.} [6] as such
\be
aa^{\dagger} - q a^{\dagger}a=q^{\alpha N + \beta },~~~
[ N, a^{\dagger} ] = a^{\dagger}, ~~ [N, a]= -a, ~~ (\alpha , \beta ~real ),
\ee
where the relation between the number operator and step operators becomes
\be
a^{\dagger} a = \cases { q^{\beta} \frac { q^{\alpha N } - q^N }{q^{\alpha } - q } & $ \alpha \ne 1 $ \cr \cr
N q^{N-1+\beta } & $ \alpha =1 $ \cr }
\ee
The forth deformation is obtained by Borzov and Damaskinsky and Yegorov [7] as such
\be
aa^{\dagger} - q^{\gamma}  a^{\dagger}a=q^{\alpha N + \beta },~~~
[ N, a^{\dagger} ] = a^{\dagger}, ~~ [N, a]= -a, ~~(\alpha , \beta , \gamma ~real ),
\ee
where the relation between the number operator and step operators becomes
\be
a^{\dagger} a = \cases { q^{\beta} \frac { q^{\alpha N } - q^{\gamma N }}{q^{\alpha } - q^{\gamma} } & $ \alpha \ne \gamma $ \cr \cr
N q^{N-1+\beta } & $ \alpha =\gamma $ \cr }
\ee
The last deformation is different from the above four cases in that it does not include the parameter $q$. Instead it has two real parameter $\alpha, \beta $ and is given by
\be
[ a, \ad ] = 2 \alpha N + \alpha + \beta, ~~~( \alpha \ge 0 , ~~ \beta >0 )
\ee

This deformed algebra is shown to be related to many interesting physical models. For $ \alpha =1, \beta = k + k' $, the spectrum is the same as the one given in the quantum system with Poeschl-Teller potential [16] whose Hamiltonian is given by
\be
\hh = \ph^2 + \frac {1}{4} \left( \frac {k(k-1)}{ \sin^2 (x/2) } +
\frac {k'(k'-1)}{ \cos^2 (x/2) } \right) , ~~~ ( 0<x<\pi)
\ee
For $ \alpha = \frac{3 \epsilon}{2} , \beta = \alpha +1 $, the algebra (11) is shown to be related to the anharmonic oscillator system with $x^4$ potential [17].

\section{ Quantization and Lie-Hamilton equation of some quantum deformed algebras }

In this section we will discuss the quantization , the generalized uncertainty relation and Lie-Hamilton equation of some quantum deformed algebras. For this work, let us assume that $a, \ad $ and $ N$ satisfies the most general deformed algebra defined by
\be
\ad a = K(N), ~~~ [ N, \ad ] = \ad , ~~~ [ N, a ] = -a
\ee
and $K(N)$ is a smooth function of $N$. Indeed, for the ordinary oscillator algebra, we have $K(N) =N $.

Let us introduce the the eigenvector $|n> $ for the number operator $N$ as follows :
\be
N|n> = n |n>, ~~~n=0,1,2 , \cdots
\ee
From the eq.(13), we know that $a$ lowers the eigenvalue by one and $a^{\dagger} $ raises the eigenvalue by one, which induce us to construct the following representation :
\be
a|n> = p(n) |n-1> , ~~~~ \ad |n> = h(n) |n+1>
\ee
From the fact that
\be
<n|\ad a |n> = \{ p(n) \}^2 ,
\ee
we obtain  the relation such as
\be
p(n) =\sqrt{ K(n) },~~~ p(n) = h(n-1)
\ee
Thus, we have the following representation
\be
a|n> = \sqrt{ K(n) }|n-1>, ~~~\ad |n> =  \sqrt{ K(n+1) }|n+1>
\ee
If we assume the existence of the ground state defined by $ K|0> =0 $, the representation is bounded below and $ K(0)=0 $. The commutation relation between $a$ and $ \ad a $ is then given by
\be
[ a, \ad ] = K(N+1) - K(N)
\ee

In order to investigate the Lie-Hamilton equation obeying the eq.(19), we define the position operator and momentum operator in terms of the generators of the most general deformed algebra :
\be
\xh = \frac {1}{2} (  \ad +  a  ) , ~~~\ph = \frac {i}{2} ( \ad -  a )
\ee
If we define the Hamiltonian $\hh $ as
\be
H = \xh^2 + \ph^2 ,
\ee
we have
\be
\hh = \frac {1}{2} ( K(N) + K(N+1) )
\ee

The commutation relation between the position operator and momentum operator is then given by
\be
[\xh, \ph ] = \frac {i}{2} ( K(N+1) - K(N) )
\ee

The Lie-Hamiltonian equation is then as follows :
\be
[ \xh , \hh ] = \frac {1}{4} \{ K(N+2) -K(N) -K(N+1) +K(N-1) \} \xh +
\frac {i}{4} \{ K(N+2) -K(N) +K(N+1) -K(N-1) \} \ph
\ee
\be
[ \ph , \hh ] = \frac {1}{4} \{ K(N+2) -K(N) -K(N+1) +K(N-1)  \} \ph -
\frac {i}{4} \{ K(N+2) -K(N) +K(N+1) -K(N-1) \} \xh
\ee
The generalized uncertainty relation is given by 
\be 
\Delta x \Delta p \ge \frac{1}{4} ( <K(N)>^2 + <K(N+1)>^2 )
\ee

Now we discuss the quantization , the generalized uncertainty relation and the Lie-Hamilton equation for the classical oscillator algebra and some of its quantum deformation.

\vspace{1cm}
{\bf Case 1 : The classical oscillator algebra}
\vspace{1cm}

In this case we have the following relation
\be
K(N) = N
\ee
Then the commutation relation between the position operator and the momentum operator is then given by
\be
[\xh , \ph ] = \frac{i}{2}
\ee
The Lie-Hamilton equation is as follows :
\be
[\xh, \hh ]= i \ph, ~~~ [\ph, \hh ] = -i\xh ,
\ee
where the Hamiltonian is given by
\be
\hh = N+ \frac {1}{2}
\ee

\vspace{1cm}
{\bf Case 2 : The $q$-deformed oscillator algebra [3]}
\vspace{1cm}

In this case we have
\be
K(N) = [N] = \frac {q^N -1}{q-1}
\ee
The Lie-Hamilton equation is as follows :
\bd
[\xh, \hh ]= - \frac{1}{4} (1-q^2 ) q^{N-1} \xh  + \frac{i}{4} (1+q)^2 q^{N-1} \ph
\ed
\be
[\ph, \hh ]=  \frac{1}{4} (1-q^2 ) q^{N-1} \ph  - \frac{i}{4} (1+q)^2 q^{N-1} \xh ,
\ee
where the Hamiltonian is given by
\be
\hh = \frac{1}{2} ( [N] + [N+1] )
\ee
Then the commutation relation between the position operator and the momentum operator is then given by
\be
[\xh , \ph ] = \frac{i}{2} q^N
\ee
From the eq.(33), we can represent the number operator in terms of the Hamiltonian as follows :
\be
q^N = \frac {2}{1+q} \{ 1-(1-q) \hh \}
\ee
Inserting the eq.(35) into the eq.(34), we obtain the following commutation relation
\be
[\xh , \ph ] = \frac {i}{1+q} \{ 1-(1-q) \hh \}
\ee
If we replace $ \xh \rightarrow \frac{\xh}{\sqrt{1+q}} $, $ \ph \rightarrow \frac{\ph}{\sqrt{1+q}} $ in the eq.(36), we have the following commutation relation
\be
[\xh , \ph ] = i \lb  1- \frac {1-q}{1+q} \hh \rb ,
\ee
which appears in the Planck scale deformed quantization [9,10]. The concrete form of the Lie-Hamilton equation is then given by
\bd
[\xh, \hh ]=  \frac{1}{2} (1-q^{-1} ) \{ 1-(1-q) \hh \}  \xh  + \frac{i}{2} (1+q^{-1} ) \{ 1- (1-q) \hh \} \ph
\ed
\be
[\ph, \hh ]=  \frac{1}{2} (q^{-1} -1  ) \{ 1- (1-q) \hh \}  \ph  - \frac{i}{2} (1+q^{-1} ) \{ 1-(1-q)\hh \} \xh ,
\ee
The generalized uncertainty relation is then given by
\be
\Delta x \Delta p \ge \frac{1}{4(1+q)} \{ 1- (1-q) ( (\Delta x)^2 +(\Delta p)^2 ) \}
\ee
This is the same as the result given in [15].

\vspace{1cm}
{\bf Case 3 : The $q$-deformed oscillator algebra [4,5]}
\vspace{1cm}

In this case we have
\be
K(N) = q^{1-N} [N]
\ee
and the Hamiltonian is given by
\be
\hh = \frac{1}{2} ( q^{1-N} [N] + q^{-N} [N+1] )
\ee
From the eq.(41), we can represent the number operator in terms of the Hamiltonian as follows :
\be
q^N = \frac {1}{1+q} \{ (q-q^{-1}) \hh + \sqrt{ (q-q^{-1})^2 \hh^2 +q^{-1} (q+1)^2 }  \}
\ee

The Lie-Hamilton equation is then given by
\bd
[\xh, \hh ]=  \frac{q^{-1}(q-q^{-1})}{2(1+q)} \hh \xh  + \frac{i}{2} \sqrt{ (q-q^{-1})^2 \hh^2 +q^{-1} (q+1)^2 } \ph
\ed
\be
[\ph, \hh ]=  - \frac{q^{-1}(q-q^{-1})}{2(1+q)} \hh \ph  - \frac{i}{2} \sqrt{ (q-q^{-1})^2 \hh^2 +q^{-1} (q+1)^2 } \xh ,
\ee

The commutation relation between the position operator and the momentum operator is then given by
\be
[\xh , \ph ] = \frac{iq}{(1+q)^2 } \sqrt{ (q-q^{-1})^2 \hh^2 +q^{-1} (q+1)^2 }
\ee

If $ q^{-1} - q $ is very small, the generalized uncertainty relation is then given by
\be
\Delta x \Delta p \ge \frac{q^{1/2}}{2(1+q)} \left(  1+ \frac{q(q-q^{-1})^2}{2(q+1)^2} ( <\xh^4 > + <\xh^2 \ph^2 > + <\ph^2 \xh^2 > + <\ph^4 > ) \right)   
\ee

\vspace{1cm}
{\bf Case 4 : The deformed oscillator algebra with non-linear spectrum [12,13]}
\vspace{1cm}

In this case we have
\be
K(N) = \alpha N^2 + \beta N
\ee
and the Hamiltonian is given by
\be
\hh = \alpha N^2 + ( \alpha + \beta )N + \frac {\alpha + \beta }{2}
\ee
From the eq.(47), we can represent the number operator in terms of the Hamiltonian as follows :
\be
N = \frac { - ( \alpha + \beta ) + \sqrt{ \beta^2 -\alpha^2 + 4\alpha \hh } }{2 \alpha }
\ee

The Lie-Hamilton equation is then given by
\bd
[\xh, \hh ]=  \alpha \xh  + i \sqrt{ \beta^2 -\alpha^2 + 4\alpha \hh } \ph
\ed
\be
[\ph, \hh ]=  \beta  \ph  - i \sqrt{ \beta^2 -\alpha^2 + 4\alpha \hh }  \xh ,
\ee

The commutation relation between the position operator and the momentum operator is then given by
\be
[\xh , \ph ] =\frac {i}{2} \sqrt{ \beta^2 -\alpha^2 + 4\alpha \hh }
\ee

If $ \beta >> \alpha $ , the generalized uncertainty relation is then given by
\be
\Delta x \Delta p \ge \frac{\beta }{4} \left(  1- \frac {\alpha}{\beta^2 } (  (\Delta x)^2 +(\Delta p)^2 ) \right)  
\ee

\section{Conclusion}

In this paper we discussed the quantization, the generalized uncertainty relation and Lie-Hamilton equation of some quantum deformed algebras. We performed this work for two types of $q$-oscillator algebras and the non-linear deformed oscillator algebra. We introduced the position operator and momentum operator in terms of the generators of the most general deformed algebra to obtain the Lie-Hamilton equation, commutation relation between the position and momentum operator and the generalized uncertainty relation which are different from the classical ( not deformed ) oscillator algebra. Moreover, for the Arik-Coon type of $q$-oscillator, the generalized uncertainty relation was shown to have the same result as the one in the Planck scale deformed quantum mechanics.

%%%%%%%%%%%%%%%%%% References
%%%%%%%%%%%%%%%%%%%%%%%%%%%%%%%%%%%%%%%%%%%%%%%%%%%%%%%%%%%%%%%%%%%%%%%
\def\JMP #1 #2 #3 {J. Math. Phys {\bf#1},\ #2 (#3)}
\def\JP #1 #2 #3 {J. Phys. A {\bf#1},\ #2 (#3)}
\def\JPD #1 #2 #3 {J. Phys. D {\bf#1},\ #2 ( #3)}
\def\PRL #1 #2 #3 { Phys. Rev. Lett. {\bf#1},\ #2 ( #3)}
\def\PLB #1 #2 #3 { Phys. Lett. B {\bf#1},\ #2 ( #3)}
\def\PRD #1 #2 #3 { Phys. Rev. D {\bf#1},\ #2 ( #3)}

%%%%%%%%%%%%%%%%%%%%%%%%%%%%%%%%%%%%%%%%%%%%%%%%%%%%%%%%%%%%%%%%%%%%%%%

\section*{References}

[1] Jimbo, Lett.Math.Phys.{\bf 10 }, 63 ( 1985) 63 ; {\bf 11}, 247 (1986 ) .

[2] Drinfeld, Proc, intern, congress of Mathematicians ( Berkeley, 1986) 78.

[3] M.Arik and D.Coon, \JMP 17 524 1976 .

[4] A.Macfarlane, \JP 22 4581 1989 .

[5] L.Biedenharn, \JP 22 L873 1990 .

[6] W.Chung, K.Chung, S.Nam and C.Um, Phys.Lett. {\bf A 183}, 363 (1993).

[7] E.Borozov, V.Damaskinsky and S.Yegorov, q-alg/9509022.

[8] F. Jackson, Mess.Math. {\bf 38 }, 57 (1909).

[9] A.Kempf, \JPD 52 1108 1995 .

[10] A.Kempf, \PRL 85 2873 2000 .

[11] G.Veneziano, Europhys.Lett.{\bf 2}199 (1986) .

[12] M.Maggiore, \PLB 304 65 1993 .

[13] E.Witten, Phys.Today {\bf 49} (4),24 (1996) .

[14] F.Scardigi, \PLB 452 39 1999 .

[15] A.Kempf, G.Mangano and R.Mann, \PRD 52 1108 1995 .

[16] J.Antoine, J.Gazeau, P.Monceau, J.Klauder and K.Penson, J.Math.Phys. {\bf 42} 2349 (2001).

[17] A.Speliotopoulos, J.Phys.A {\bf 33}, 3809 (2000).

\end{document}